\documentclass[12pt,preprint]{aastex}

\def\parcsec{\ensuremath{\,.\!\!\arcsec}}
\def\kms{km\,s$^{-1}$}

\def\gtrsim{\mathrel{\hbox{\rlap{\hbox{\lower4pt\hbox{$\sim$}}}\hbox{$>$}}}}
\def\lesssim{\mathrel{\hbox{\rlap{\hbox{\lower4pt\hbox{$\sim$}}}\hbox{$<$}}}}
\def\HeI{He\,{\sc i}}

\def\CaII{Ca\,{\sc ii}}

\shorttitle{Progenitor of SN~2008cn}
\shortauthors{Elias-Rosa et al.}

\begin{document}



\title{On the Progenitor of the Type II-Plateau SN~2008cn in
NGC~4603\footnote{Based on observations made with the NASA/ESA {\it Hubble
Space Telescope}, obtained from the Data Archive at the Space Telescope
Science Institute, which is operated by the Association of
Universities for Research in Astronomy, Inc., under NASA contract NAS
05-26555. Based in part on data gathered with the 6.5-m Magellan
telescopes, located at Las Campanas Observatory, Chile.}}


\author{Nancy Elias-Rosa\altaffilmark{2},
  Schuyler D.~Van Dyk\altaffilmark{2},
  Weidong Li\altaffilmark{3},
  Nidia Morrell\altaffilmark{4},
  Sergio Gonzalez\altaffilmark{4},
  Mario Hamuy\altaffilmark{5},
  Alexei V. Filippenko\altaffilmark{3},
  Jean-Charles Cuillandre\altaffilmark{6},
  Ryan J. Foley\altaffilmark{3,7,8}, and
  Nathan Smith\altaffilmark{3}
}

\altaffiltext{2}{Spitzer Science Center, California Institute of 
Technology, 1200 E. California Blvd., Pasadena, CA 91125; email 
nelias@ipac.caltech.edu, vandyk@ipac.caltech.edu .}
\altaffiltext{3}{Department of Astronomy, University of California,
Berkeley, CA 94720-3411.}
\altaffiltext{4}{Las Campanas Observatory, Carnegie Observatories, 
Casilla 601, La Serena, Chile.}
\altaffiltext{5}{Departamento de Astronomia, Universidad de Chile, 
Casilla 36-D, Santiago, Chile.}
\altaffiltext{6}{Canada-France-Hawaii Telescope Corporation, 65-1238 
Mamalahoa Hwy, Kamuela, HI 96743.}
\altaffiltext{7}{Harvard/Smithsonian Center for Astrophysics, 60 
Garden Street, Cambridge, MA 02138.}
\altaffiltext{8}{Clay Fellow.}




\begin{abstract}
A trend is emerging regarding the progenitor stars that give rise to
the most common core-collapse supernovae (SNe), those of Type
II-Plateau (II-P): they generally appear to be red supergiants with a
limited range of initial masses, $\sim$8--16 M$_{\sun}$. Here we
consider another example, SN 2008cn, in the nearly face-on spiral
galaxy NGC~4603.  Even with limited photometric data, it appears that
SN 2008cn is not a normal SN II-P, but is of the high-luminosity
subclass. Through comparison of pre- and post-explosion images
obtained with the Wide Field and Planetary Camera 2 (WFPC2) onboard
the {\sl Hubble Space Telescope (HST)}, we have isolated a supergiant
star prior to explosion at nearly the same position as the SN.  We
provide evidence that this supergiant may well be the progenitor of
the SN, although this identification is not entirely unambiguous.
This is exacerbated by the distance to the host galaxy, 33.3 Mpc,
making SN 2008cn the most distant SN II-P yet for which an attempt has
been made to identify a progenitor star in pre-SN images.  The
progenitor candidate has a more yellow color ($[V-I]_0=0.98$ mag and
$T_{\rm eff} = 5200 \pm 300$ K) than generally would be expected and,
if a single star, would require that it exploded during a ``blue
loop'' evolutionary phase, which is theoretically not expected to
occur. Nonetheless, we estimate an initial mass of
$M_{\rm ini} = 15 \pm 2\ {\rm M}_{\sun}$ for this star, which is
within the expected mass range for SN~II-P progenitors. The yellower
color could also arise from the blend of two or more stars, such as a
red supergiant and a brighter, blue supergiant. Such a red supergiant
hidden in this blend could instead be the progenitor and would also
have an initial mass within the expected progenitor mass
range. Furthermore, the yellow supergiant could be in a massive,
interacting binary system, analogous to the possible yellow supergiant
progenitor of the high-luminosity SN II-P 2004et. Finally, if the
yellow supergiant is not the progenitor, or is not a stellar blend or
binary containing the progenitor, then we constrain any undetected
progenitor star to be a red supergiant with $ M_{\rm ini} \lesssim
11\ {\rm M}_{\sun}$, considering a physically more realistic scenario
of explosion at the model endpoint luminosity for a rotating star.
\end{abstract}


\section{Introduction}\label{introduction}

Massive stars are thought to evolve to an end state which results in
the collapse of the stellar core, as the hydrostatic pressure in the
core can no longer support gravity.  Although the physics of what
follows is still uncertain, it appears that, following core collapse,
the pressure due to neutrinos released from the core results in the
explosion of the rest of the star --- a supernova (SN; e.g., 
\citealt{marek09}).
Core-collapse supernovae (CC-SNe) have been at the forefront of
astronomical research for the better part of a century, and yet we
still do not have complete knowledge of which stars explode. It is
possible that the relative heterogeneity in observed properties of
CC-SNe arises from a range of massive progenitors.  Although indirect
evidence (spectral features, light-curve shapes, environmental
information) provides some clues, direct identification of the star
before the SN explosion \citep[e.g.,][]{vandyk99,vandyk00,smartt02} is
the more precise means of determining the nature of the progenitor star.

At this time of writing, the only kind of CC-SNe for which progenitors
have been directly identified is the Type II-Plateau supernovae 
(SNe~II-P; e.g.,
\citealt{vandyk03,li06}), with the exception of SN~2005gl in NGC~266,
a Type II-narrow supernova (SN~IIn) for which a progenitor may well have been
identified \citep{galyam07,galyam09}. The SNe~II-P exhibit Balmer lines
with prominent P~Cygni-like profiles in their optical spectra at early
times.  Their optical light-curve shape shows a characteristic
``plateau'' phase, prior to an exponential tail (e.g.,
\citealt{barbon79}).  The trend is for their progenitor masses to be in the range of
8--20 $M_{\odot}$ (\citealt{li06}).
Based on the statistics of 20 SNe~II-P for which
progenitors have been isolated or upper mass limits established,
\citet{smartt09} derive a more limited range of $8.5 (^{+1}_{-1.5}) \lesssim M_{\rm
ZAMS} (M_{\odot}) \lesssim 16.5 (\pm 1.5)$ for these stars. All
of these progenitors exploded in the red supergiant (RSG) phase, as we
would theoretically expect, except possibly for SN 2004et \citep{li05} and SN 2006ov \citep{li07}. 
Given that fewer than 10 progenitors for SNe II-P have
actually been directly identified, the addition of further
examples is essential for validating this trend and further
constraining stellar evolutionary theory.

The explosion of \object{SN~2008cn} provides us with an
opportunity to increase the sample, including particularly the high-luminosity SN subclass (e.g., \citealt{pastorello03}), and to extend our techniques to the study of stars which have exploded in relatively more distant galaxies. 

\citet{martin08} discovered SN~2008cn on 2008 May 21.52 (UT dates are
used throughout this paper), at 4$\parcsec$7 E and 23$\parcsec$2 N of
the center of the nearby ($cz = 2592$ \kms), nearly face-on spiral
galaxy \object{NGC~4603}.  Nine days later, \citet{strit08} classified
the SN as Type II, a few days past explosion.  \citet{li08} measured a
precise position for a putative progenitor of SN~2008cn as
$\alpha=12^{\rm h} 40^{\rm m} 55{\fs}64, \delta=-40\arcdeg 58\arcmin
12{\farcs}9$ (J2000.0), from high-quality {\sl Hubble Space Telescope\/} 
({\sl HST}) archival images of the host galaxy taken
with the Wide Field Planetary Camera~2 (WFPC2).  The position was
established after the {\sl HST\/} images were registered with coadded
$V$-band images taken with the 1.0-m Swope telescope at Las Campanas
Observatory on 2008 June 1.15 (a geometrical transformation was
performed, resulting in an accuracy of $0{\farcs}1$ on the WFPC2
images). The claim was that the detected object could be a RSG with an
absolute $I$-band magnitude of $-7.9$ and an intrinsic $V-I$ color of
1.2 mag.

Here we report on a more detailed analysis of this object.  In \S \ref{specph} we determine the characteristics of SN~2008cn. Section
\ref{identification} explains the process of identification and
analysis of the possible progenitor using {\sl HST\/} images taken both 
before, and after, the SN explosion. We discuss the nature of the 
progenitor, based on this analysis, in \S \ref{discussion}, and in \S
\ref{conclusions} we summarize our conclusions.

\section{The Nature of SN~2008cn}\label{specph}

Before discussing the nature of the progenitor of  SN~2008cn, we should derive more information from observations of the SN itself (its spectrum and light curves). Unfortunately, we have only scant coverage, but even from this partial dataset we are able to produce a relatively consistent
reddening estimate and can characterize SN~2008cn.

In addition to the images obtained with the Swope 1.0-m telescope on
2008 June 1, $V$-band images of the SN were also obtained on 2008 June
15 with the Baade 6.5-m Magellan telescope + the Inamori-Magellan
Areal Camera \& Spectrograph (IMACS) at Las Campanas Observatory; see
Table \ref{table_observ}. The images were trimmed, corrected for
overscan and bias, and flat-fielded. Instrumental magnitudes for the
SN and the stars in its immediate environment were measured using
SNOoPY (SuperNOvaPhotometrY)\footnote[9]{SNOoPY is a package originally
  designed by F. Patat and later implemented in IRAF by E.
  Cappellaro.}, which is based on DAOPHOT \citep{stetson87} within
IRAF\footnote[10]{IRAF (Image Reduction and Analysis Facility) is
  distributed by the National Optical Astronomy Observatories, which
  are operated by the Association of Universities for Research in
  Astronomy, Inc., under cooperative agreement with the National
  Science Foundation.} and allows for fitting of the stellar profiles
with a point-spread function (PSF) created from a set of unsaturated
stars in the images. The SN magnitudes were calibrated using the
comparison stars in the images, based on four other photometric nights
when the SN was observed (C.~Contreras, private communication).  We
also obtained late-time photometry for the SN from our {\sl HST\/}
Target of Opportunity (ToO) observations (see \S \ref{identification})
and a late-time upper limit ($\sim$ 416 days after explosion) from
images obtained on 2009 July 01.9 with the Cerro Tololo
InterAmerican Observatory (CTIO) 1.3-m telescope + ANDICAM, operated
by the SMARTS consortium. The SN photometry is listed in Table
\ref{table_ph}.

We have also analyzed an unpublished high-resolution optical spectrum
of the SN, obtained on 2008 June 16 with the Baade 6.5-m Magellan
telescope + IMACS. The complete set of ground-based observations,
including this spectrum, will be presented elsewhere by the Carnegie
Supernova Project
[CSP]\footnote[11]{http://csp1.lco.cl/~cspuser1/PUB/CSP.html .}.  The
spectrum appears quite similar to that of the high-luminosity SN~1996W
\citep{pastorello03,inserra09} at $\sim$
1 month after explosion. The SN~2008cn spectrum exhibits similar P
Cygni profiles for the H Balmer emission lines; however, it lacks
\HeI\ $\lambda$5876, which is seen in the SN 1996W spectrum at this
phase. We derive an estimate of the explosion date (JD 2454598 $\pm\ $
2), which is consistent with the date of the upper limit to the SN
detection (May 8.8) given by \citet{martin08}. SN~2008cn also presents
similarities with the spectra of other high-luminosity events, such as
SNe 1992H \citep{clocchiatti96} and 2003hn \citep{krisciunas09} at
coeval epochs, even if these last two show a more evolved
\CaII\ near-infrared triplet (\CaII\ $\lambda\lambda$8498, 8542,
8662). In general, as has been discussed by \citet{pastorello03}, all
of these high-luminosity SN spectra show a slower evolution, compared
with those of intermediate luminosity, such as SN~1999em
\citep{hamuy01,leonard02}.

Next, we estimated the reddening to SN 2008cn from the spectrum.
Measuring the equivalent width (EW) of the Na\,{\sc i} D line at the
host-galaxy redshift ($z=0.009$) in the spectrum, we find that
EW(Na\,{\sc i} D) = 1.7 \AA. Using the relation between extinction and
EW(Na\,{\sc i} D) from \citet{elias09}, and
assuming the \citet{cardelli89} reddening law with updated wavelengths
(as discussed by \citealt{elias09}) and a
Galactic foreground $E(V-I)$ = 0.22 mag \citep{schlegel98}, we
derive ${E(V-I)}_{\rm tot} = 0.44 \pm 0.06$ mag (${E[B-V]}_{\rm tot} =
0.35 \pm 0.04$ mag). We adopt this value for SN~2008cn.

\begin{figure*}
\epsscale{0.5}
\centering
\includegraphics[height=5truein,width=3.8truein,angle=-90]{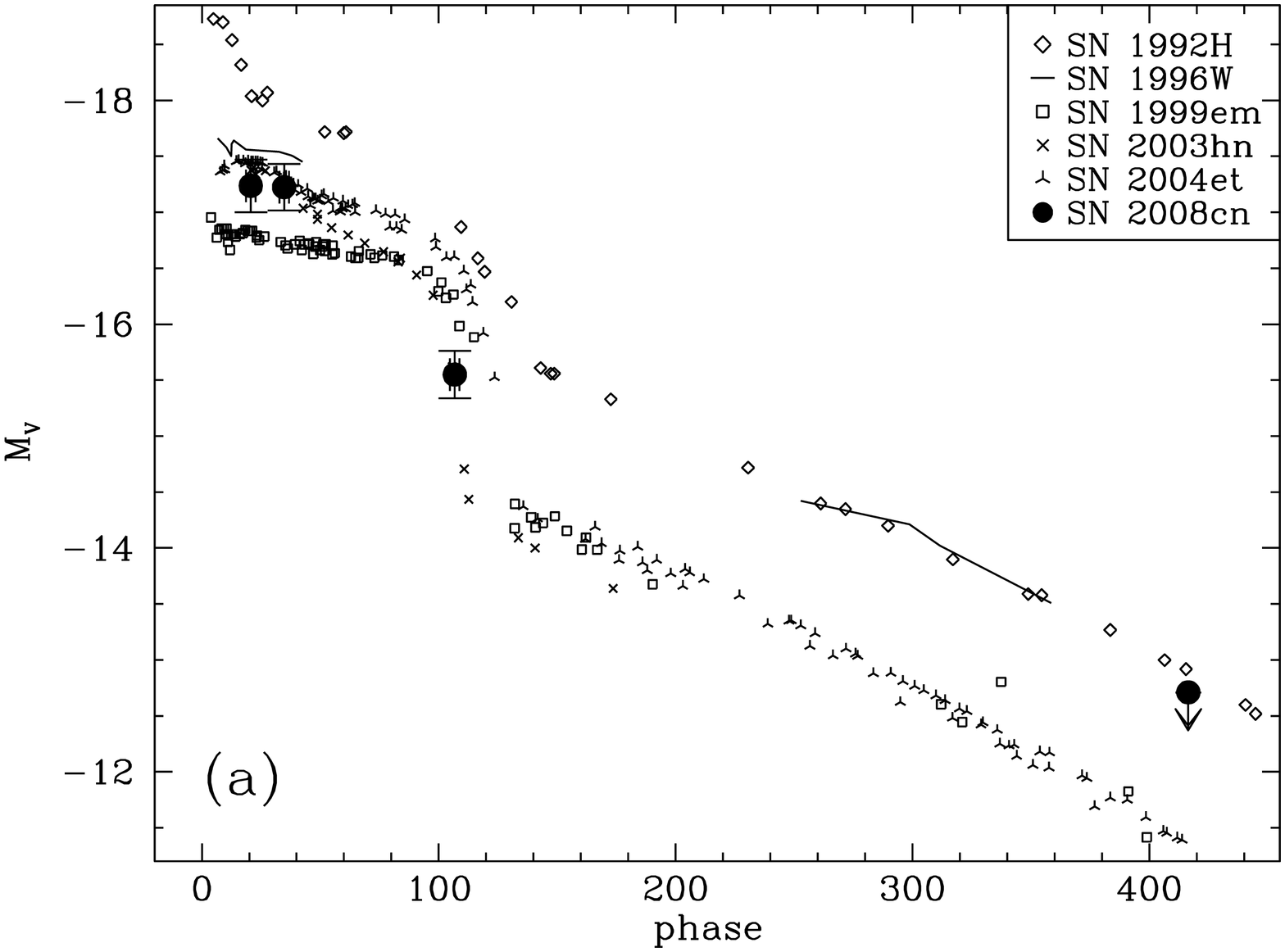}
\includegraphics[height=5truein,width=3.8truein,angle=-90]{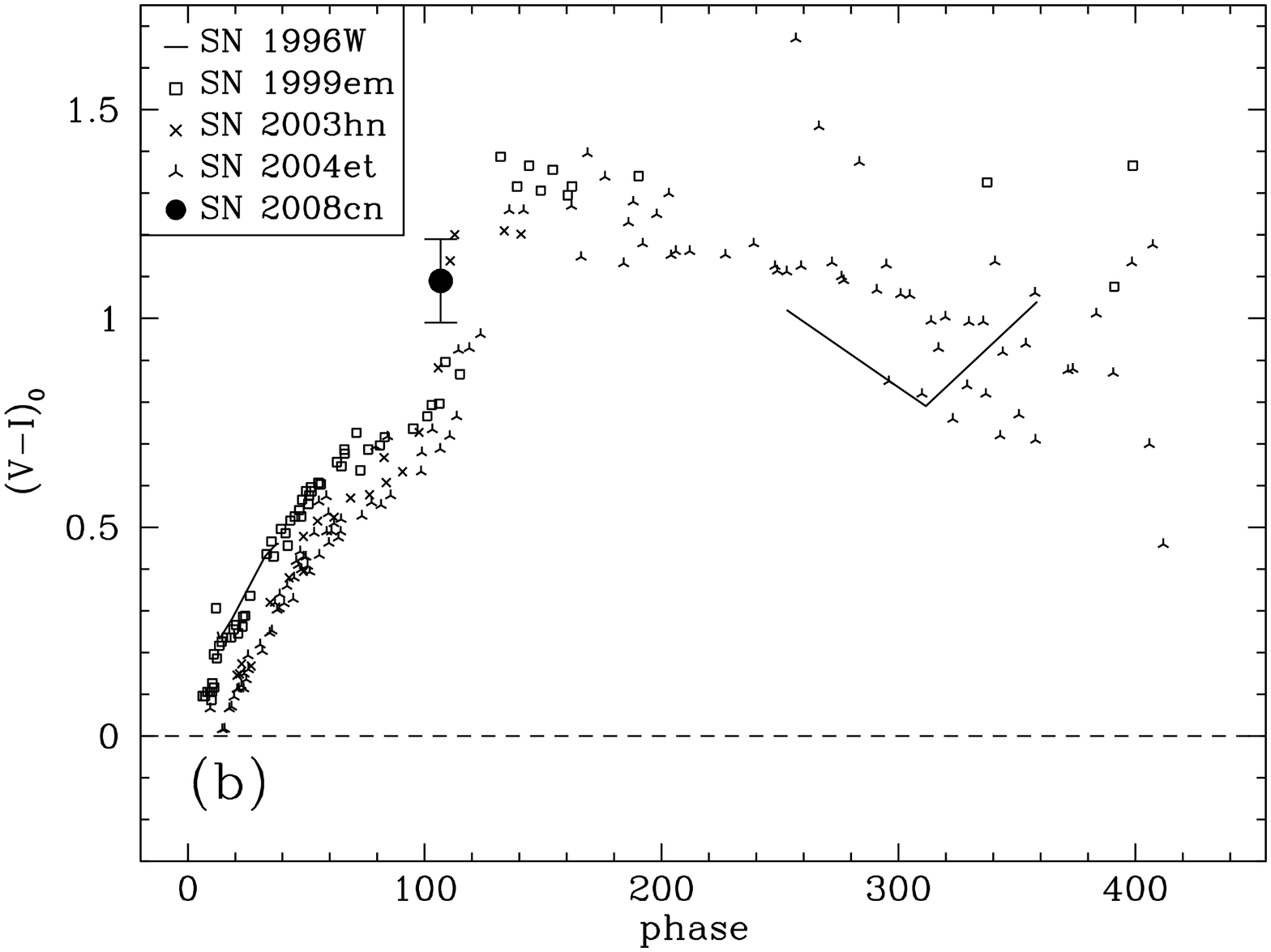}
\caption{{\it Top - (a):} Absolute $V$ light curve of SN~2008cn 
({\it filled circle}) along with those of the high-luminosity SNe II-P
  1992H ({\it diamonds}), 1996W ({\it solid line}), 2003hn 
  ({\it crosses}), and 2004et ({\it pentagons}), 
  as well as of the normal SN II-P~1999em
  ({\it squares}). Distances and extinction estimates for these latter
  SNe have been adopted from the literature. {\it Bottom - (b):} The
  intrinsic $(V-I)_0$ color of SN~2008cn, compared with the intrinsic
  color evolution of the latter SNe. The color of SN~2008cn has been
  corrected for the assumed reddening, $E(V-I)$ = 0.44 mag. See text
  for additional details.
\label{fig_color}}
\end{figure*}

Adopting the distance to NGC~4603 determined by \citet{newman99} using
Cepheid variables, 33.3$\pm$0.2 Mpc, together with our extinction
estimate and the explosion date suggested by the spectrum, we have
compared our $V$ absolute light curve, however limited, with those of
SNe~1992H \citep{clocchiatti96}, 1996W \citep{pastorello03,inserra09},
1999em \citep{hamuy01,leonard02}, 2003hn \citep{krisciunas09}, and
2004et \citep{sahu06}; see Figure \ref{fig_color} {\it top}. 
All of these latter SNe are likely of high luminosity,
except for SN~1999em. The overall shape of the light curves of
SNe~2008cn and 2003hn are very similar, and the length of the plateau
for each is also almost the same. SN~2008cn is less luminous than the
other high-luminosity SNe II-P, but it is clearly more luminous
(by $\sim$ 0.5 mag) than SN~1999em. The mean absolute $V$ magnitude
during the plateau phase, estimated by averaging the magnitudes from
20 to 35 days after explosion, is $-$17.23, brighter than $\sim$ 90\%
of the SNe II-P analyzed in a complete sample by Li et al. (2009, in
preparation). From our data, the SN~2008cn plateau may well be as
steep as that of SN~2003hn. Unfortunately, we were not able to
compare properly the SN 2008cn light curve with that of SN~1996W,
given the lack of observations for this latter SN at the end of the
plateau phase.

In Figure \ref{fig_color} ({\it bottom}) we can also see the
comparison of the intrinsic $(V-I)_0$ color for SN~2008cn, from our
late-time {\sl HST\/} observations (see \S \ref{identification}), to
$(V-I)_0$ for the other SNe considered above. These latter SN colors
have been corrected for extinction using published estimates and
assuming the \citet{cardelli89} extinction law. In the case of
SN~2008cn, we have corrected the observed color by the reddening
discussed above.

Both the spectroscopy and the photometry of SN~2008cn indicate that
it is a highly luminous SN~II-P.  For only one other SN of this kind,
SN~2004et, has a progenitor star been identified (\citealt{li05};
but see \citealt{smartt09}). As discussed by
\citet{pastorello03}, semi-analytical models predict that these SNe
should have massive progenitors, near the high end of the mass range
for SNe II-P determined by \citet{smartt09}. Thus, locating the
progenitor of SN 2008cn could enrich our knowledge of the progenitor
stars for these unusual SNe.


\section{Identification and Analysis of the Progenitor Candidate}\label{identification}

As can be seen in the next subsections, the analysis of the location
of the possible progenitor of SN~2008cn was developed following
several steps: (1) search the {\sl HST\/} archive for pre-SN images of
the SN; (2) locate the progenitor candidate in these pre-explosion
images, assisted by ground-based images of the SN at early times; and,
(3) confirm the candidate, first identified by \citet{li08}, using
post-explosion {\sl HST\/} images obtained as part of our ToO program GO-11119.


\subsection{Isolating the Progenitor Candidate}\label{low_identification}

Once SN~2008cn was discovered, we found 48 images in the F555W ($\sim
V$) and 12 images in F814W ($\sim I$) filters in the {\sl HST\/}
archive\footnote[12]{http://archive.stsci.edu/hst/.}. These images were
all obtained with WFPC2 between 1996 and 1997 by program GO-6439 (PI:
S.~Zepf) in order to determine the distance to NGC 4603 using Cepheid
variables (Newman et al. 1999). At each epoch, images in each band
were obtained in cosmic-ray split (CR-SPLIT) pairs; see Table
\ref{table_observ} for a complete list of the pre-explosion
observations analyzed here.  In all of these images the site of
SN~2008cn was located on the WF2 chip (pixel scale of $0{\farcs}1$
pix$^{-1}$). The exposure times in each band were in the range of
900--1300~s. Combining the images in each band over all epochs of the
Cepheid observations yields a total image depth of $\sim$16~hr in
F555W and $\sim$ 4~hr in F814W.

\begin{figure*}
\epsscale{1.0}
\plotone{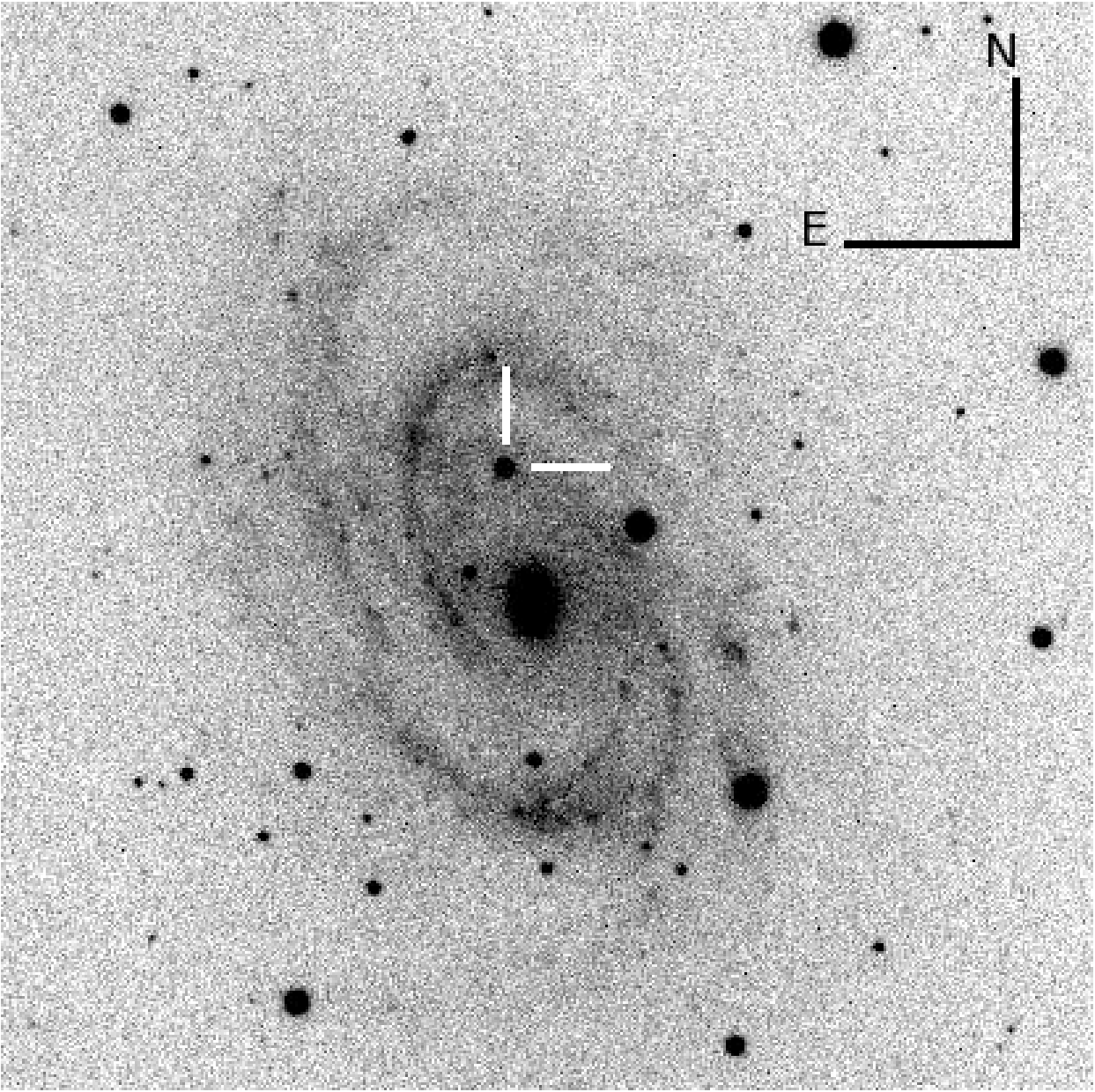}
\caption{Baade 6.5-m Magellan Telescope + IMACS
$V$-band image of SN~2008cn. The SN is indicated by the 
white arrows. The field of view is $\sim 4\arcmin \times 4\arcmin$.
\label{fig_magimage}}
\end{figure*}

To obtain a more accurate position of the SN~2008cn progenitor
candidate in the archival {\sl HST\/} images than what \citet{li08}
were able to accomplish with the 1.0-m images, we compared the {\sl
  HST\/} images with the Magellan $V$-band image of the SN (with pixel
scale $0{\farcs}28$ pix$^{-1}$) from 2008 June 15; see Figure
\ref{fig_magimage}. We combined two of the CR-SPLIT {\sl HST\/}
exposures in F555W (images u38l0101t, u38l0102t) in order to reject
cosmic-ray hits, and produced a $1600 \times 1600$ pixel mosaic of all
four chips from the combined image, using the routines {\it crrej} and
{\it wmosaic} of the STSDAS package within IRAF. We then used the task
{\it rotate\/} within IRAF to rotate the mosaic around its center
(pixel 800, 800) by the angle defined in the mosaic header keyword
ORIENTAT.  We identified 7--11 point-like sources in common between
the mosaic and the Magellan image and measured their pixel coordinates
with {\it imexamine}. Using the IRAF task {\it geomap}, we carried out
a geometrical transformation between the two sets of coordinates, with
root-mean square (rms) uncertainty $\lesssim 0{\farcs}1$.  With the
IRAF task {\it geotran\/} we registered the Magellan $V$ image to the
coordinate frame of the rotated {\sl HST\/} F555W mosaic, which
allowed us to measure a precise pixel position for the SN in this
latter image at [1132.58, 843.99].  Registering available point
sources to the reference frame of the Two Micron All Sky Survey
(2MASS\footnote[13]{http://www.ipac.caltech.edu/2mass/.}) Point Source
Catalog, this pixel position corresponds to $\alpha=12^{\rm h} 40^{\rm
  m} 55{\fs}64, \delta=-40\arcdeg 58\arcmin 12{\farcs}1$ (J2000.0).
Considering the 2MASS positional rms uncertainty of $\sim$ 0\parcsec2,
the total uncertainties estimated in this position are $\Delta \alpha$
= 0\parcsec2 and $\Delta \delta$ = 0\parcsec3.

Through careful examination of the image mosaic, the object nearest
the SN site, as determined from images obtained in both {\sl HST\/}
filters, is located at $\alpha=12^{\rm h} 40^{\rm m} 55{\fs}63,
\delta=-40\arcdeg 58\arcmin 12{\farcs}2$ (pixel [1134.37, 843.00]). We
consider this object, the same as that suggested by \citet{li08}, to
be the progenitor candidate.  Given that the offset between the SN
position (measured from the ground-based image) and the candidate
position is significant ($\Delta x = 1.79$, $\Delta y = 0.99$), it
became essential that we use even higher-resolution images of the SN
(ideally, at the same resolution as the pre-SN {\sl HST\/} image) to
attempt to reduce this offset, before further considering this
candidate as the actual progenitor star.


\subsection{Confirmation of the Progenitor Location}\label{high_identification}

Post-explosion ToO {\sl HST\/} WFPC2 observations were obtained in
F555W and F814W on 2008 August 26 (see Table \ref{table_observ} for
more details) as part of our program GO-11119. The SN was imaged on
the PC chip (pixel scale $0{\farcs}045$ pix$^{-1}$).  The SN was still
quite bright at the time of these observations, so we used the gain
setting of 14 e$^-$DN$^{-1}$.  Furthermore, we obtained a set of images in each band
with short exposure times, with no dithering between images, and each
set was combined into a single image per band (see below).

\begin{figure*}
\epsscale{1.0} 
\plotone{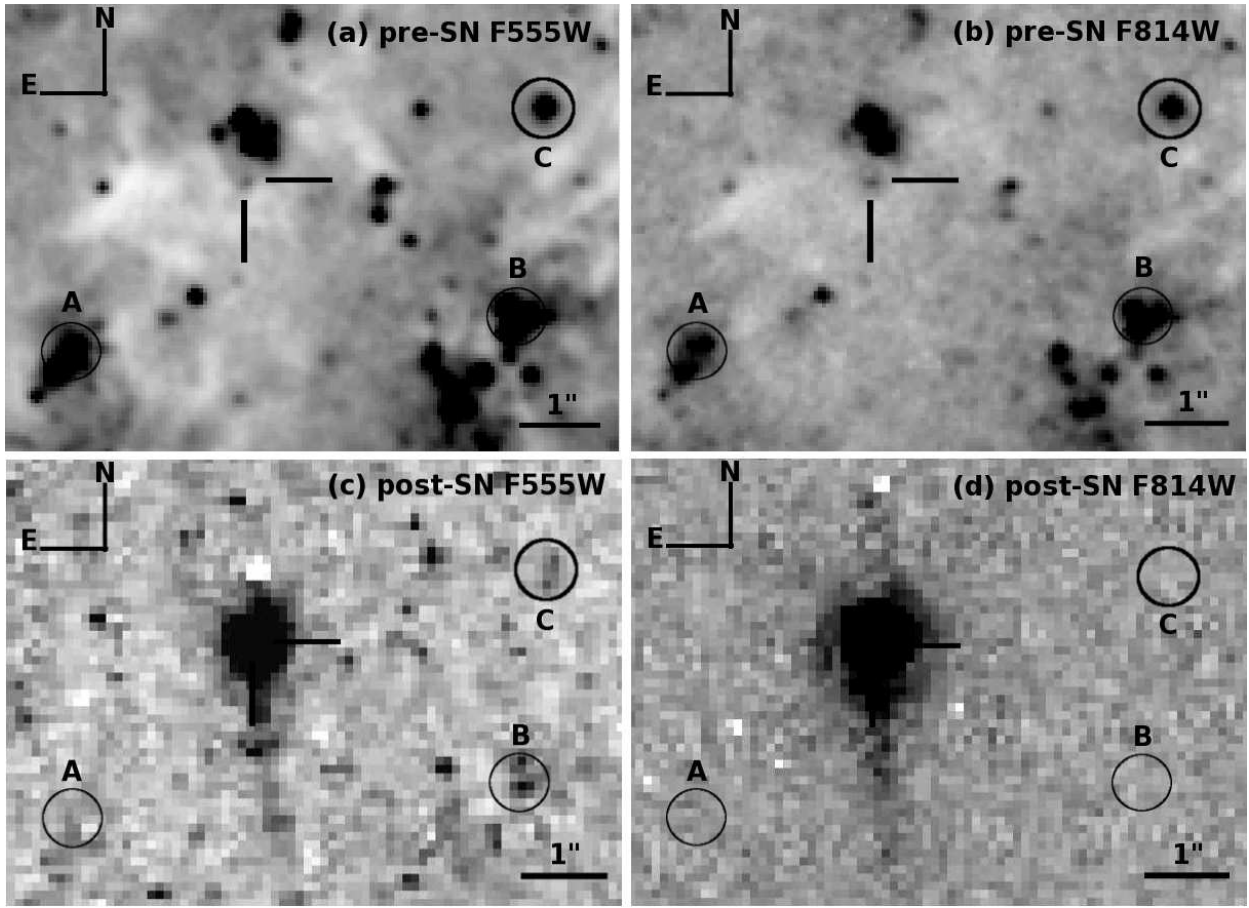}
\caption{Subsections of the {\sl HST\/} + WFPC2 images, in F555W 
  ({\it left}) and F814W ({\it right}), of NGC~4603 before 
  ({\it panels a\/} and {\it b}) and after ({\it panels c} and {\it d}) the
  SN~2008cn explosion. The positions of the SN candidate progenitor and
  SN are indicated by {\it bars}.  The three objects ``A,'' ``B,'' and
  ``C'' were not used in the image astrometry, due to their relative
  faintness; they are labeled only to aid the eye in locating the
  progenitor in the field.}
\label{fig_progenitor}
\end{figure*}

To achieve high-precision relative astrometry between the SN and
progenitor candidate positions, we needed to geometrically transform
these post-explosion images to match the pre-explosion ones.
Pre-explosion image mosaics for each band, which have been combined
from the individual exposures (see \S~\ref{low_identification}) to the
pixel scale of the PC chip using the {\it drizzle\/} algorithm
\citep{fruchter02}, are available from the {\sl HST\/} Legacy
Archive\footnote[14]{http://archive.stsci.edu/hlsp/.}.  The {\it drizzle\/}
image processing method weights input images according to the
statistical significance of each pixel, and removes the effects of
geometric distortion on both image shape and photometry. In addition,
it combines dithered images in the presence of cosmic rays and
improves the resolution of the mosaic. Similarly, we ``drizzled'' the
post-explosion images to the PC chip resolution, following the recipe
provided by the Space Telescope Science
Institute\footnote[15]{http://www.stsci.edu/hst/wfpc2/analysis/drizzle.html.}. 
The task {\it drizzle\/} runs under PYRAF as part of the STSDAS
package. Next, the registration of the pre-explosion and
post-explosion image mosaics was performed using eleven point-like
sources in common between the two sets of data. The number of
available fiducial stars is limited by the relatively shallow depth of
the combined ToO exposures. The fiducial sources are faint; however,
they are concentrated around the SN site, reducing the impact of any
possible residual distortion remaining in the larger
mosaics. Comparing pixel-by-pixel both sets of data, we verified the
proximity of the point-like progenitor candidate in the pre-explosion
images very near the location of the SN; see Figure
\ref{fig_progenitor}. Using two methods, the task {\it daofind\/}
within IRAF/DAOPHOT and {\it imexamine}, we measured the positions of
both the SN and progenitor candidate. We averaged the results from
both methods to establish final pixel positions, which are [2458.13,
  2045.74] and [2457.86, 2045.22] for the SN and [2457.94, 2046.54]
and [2457.96, 2045.89] for the progenitor candidate, in F555W and
F814W, respectively. Note that no other source was located within a
1.5$\sigma$ ($\leq$2 pixels) radius from the progenitor candidate
position. The differences between the SN and the progenitor candidate
positions, compared with the total estimated uncertainty in the
astrometry, are given in Table \ref{table_error}.  This latter
uncertainty was calculated as a quadrature sum of the uncertainties in
the SN and progenitor candidate positions, and the rms uncertainty in
the geometric transformation.

From the results in Table \ref{table_error} it can be seen that the difference between the SN position and the position of the progenitor candidate in declination is somewhat larger than the total astrometric error, for both bands. Although the relative positional coincidence between the
candidate and SN position in right ascension suggests that we
have indeed identified the SN progenitor star, the small offset in
declination somewhat weakens our confidence.  We considered it
possible that the presence of the two, much brighter point sources
within $\sim 1\arcsec$ to the north of the candidate (Figure
\ref{fig_progenitor}, panels $a$ and $b$) may be affecting the
measurement of the centroid in declination, resulting in a northward
skew in the progenitor candidate position using both of our methods.  For this reason, we
repeated the measurements described above after first masking out
these two bright stars in the pre-explosion images.  Doing so,
however, did not significantly reduce the positional difference in
declination between the candidate and the SN in either band.

For now, we make the assumption that the candidate is the
progenitor star of SN 2008cn and discuss below the implications of
this assumption.  However, given the slight positional difference, 
we will also subsequently consider the implications if this star is
{\it not\/} the SN progenitor.


\subsection{Photometry of the Pre- and Post-Explosion {\sl HST\/} Images}\label{ph}

We obtained photometry of both the progenitor candidate in the
pre-explosion {\sl HST\/} images and the SN in the post-explosion {\sl
HST\/} images.  Photometry of all of the WFPC2 images was performed
using the package HSTphot\footnote[16]{HSTphot is a stellar
photometry package specifically designed for use with {\sl HST\/}
WFPC2 images. We used v1.1.7b, updated 19 July 2008.}
(\citealt{dolphin00a}). Since our pre-explosion images were obtained
at different epochs and at various pointings, and not always in
matched pairs of bands, we measured the relative offsets between the
pre-explosion images (the orientation angle for all of these {\sl HST\/}
images is $143{\fdg}89$). Once corrected for relative offsets
with respect to one fiducial image, we ran the package on the CR-SPLIT
pairs of the individual frames in each band separately, with option
flag 10, which combines turning on the local sky determination, turning off
empirically determined aperture corrections (using default values
instead), and turning on PSF residual
determination, with a total detection threshold of $3\sigma$. We then
transformed the flight-system magnitudes in F555W and F814W to the
corresponding Johnson-Cousins \citep{bessell90} magnitudes (in $V$ and $I$),
following the prescriptions of \citet{holtzman95} with updates
from \citet{dolphin00b}.

Since the post-explosion images were all obtained at the same epoch in
each band and without any dithering between exposures, we
straightforwardly input all of the individual exposures into HSTphot.
The output from the package automatically includes the transformation
from flight-system F555W and F814W to Johnson-Cousins $V$ and $I$.

Table \ref{table_ph} lists the photometry estimated for the SN and
for the progenitor candidate.


\section{The Progenitor of SN~2008cn}\label{discussion}


To further analyze the progenitor candidate, we must convert its
observed brightness and color into intrinsic properties, such as
luminosity and effective temperature.  Doing so, we can attempt to
place the star on a Hertzsprung-Russell (HR) diagram and draw some
conclusions about its evolutionary state and initial mass.

From our adopted values for the distance and extinction to the SN, we
find that the absolute magnitudes of the progenitor candidate are
$M_V^0=-7.31 \pm 0.21$ and $M_I^0=-8.29 \pm 0.23$. These values and
the resulting intrinsic color, $(V-I)_0=0.98$ mag, are more consistent
with a supergiant star of spectral type G than a much cooler RSG. If
this star is the progenitor, it is far more yellow than the RSG
progenitors so far identified, which would be an unusual, although not
necessarily unprecedented, result. We note that \citet{li05} isolated
a yellow supergiant as the progenitor of the high-luminosity SN II-P
2004et in NGC~6946 (but see \citealt{smartt09}). SN 2004et
appears to have been somewhat unusual in its photometric and
spectroscopic properties \citep{li05,sahu06}. Furthermore, the locus
of the progenitor of SN 2006ov in M61 on the HR diagram may also
indicate that the star was more yellow than red \citep[][although,
  again, see \citealt{smartt09}]{li07,prieto08}.

The metallicity at the SN site is likely solar. We determine this to
be the case, based on the position of the SN in the host galaxy and
the result derived from the typical central metallicity (12 + log[O/H]
= 8.99 $\pm$ 0.24) and gradient ($-0.06$ dex kpc$^{-1}$) for a sample
of 24 late-type spiral galaxies (\citealt{zaritsky94}). Thus, for the
projected distance from the center of NGC~4603 to SN~2008cn (3.8 kpc), we
estimate 12 + log(O/H) = 8.76 $\pm$ 0.24, which is comparable to the
solar value (12 + log[O/H] = 8.66 $\pm$ 0.05; \citealt{asplund05}).

For solar metallicity, the derived color for the star corresponds to
an effective temperature $T_{\rm eff} = 5200 \pm 300$ K (the
uncertainty is a conservative estimate). For an assumed surface
gravity $\log g = +0.5$, according to the Kurucz Atlas 9 models
(\citealt{kurucz93}), the bolometric correction to the absolute
magnitude is $-0.29$ mag. Thus, the progenitor candidate has an
absolute bolometric luminosity $L_{\rm bol}=10^{(4.93 \pm
  0.10)}\ L_{\sun}$ (assuming the Sun's absolute bolometric magnitude
is 4.74).

\begin{figure*}
\epsscale{1.0}
\centering
\includegraphics[height=5truein,width=3.8truein,angle=-90]{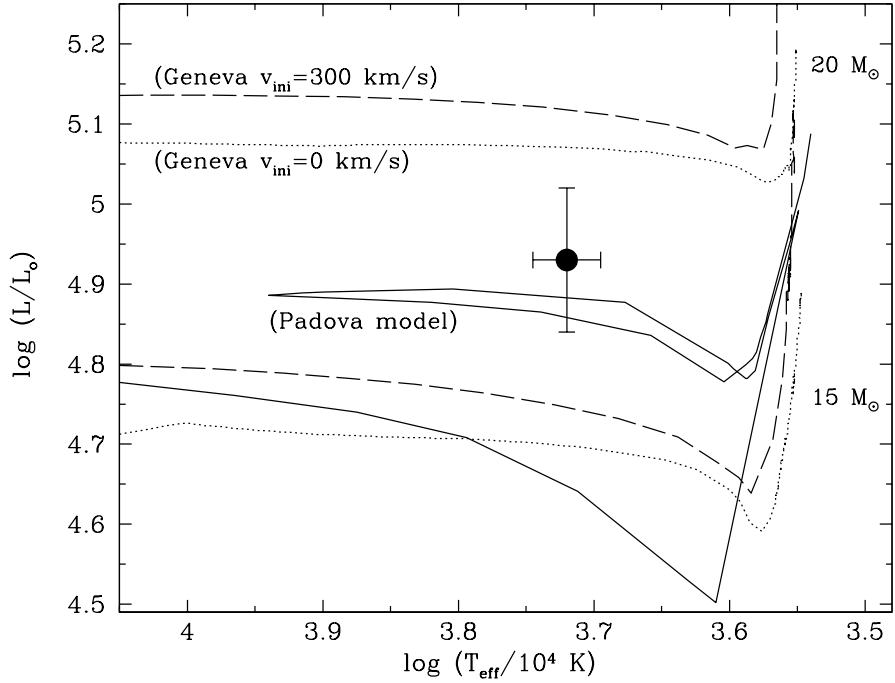}
\caption{A Hertzsprung-Russell diagram showing the bolometric
  luminosity, $L_{\rm bol}$, and effective temperature, $T_{\rm eff}$,
  for the candidate progenitor of SN~2008cn ({\it filled circle}).  We
  have assumed a reddening $E(V-I)_{\rm tot} = 0.44$ mag and a
  distance modulus $\mu_0 = 32.61 \pm 0.20$ mag \citep{newman99} for
  the star. Model stellar evolutionary tracks for a solar metallicity
  ($Z = 0.02$) are also shown for a range of masses and a rotation of
  $v_{\rm ini} = 0$ \kms (i.e., no rotation; {\it solid line} for the
  15 M$_{\sun}$ track from \citealt{bressan93} and {\it dotted lines}
  for the tracks from \citealt{hirschi04}) and $v_{\rm ini} = 300$
  \kms\ ({\it dashed lines}, \citealt{hirschi04}).
\label{fig_hrd}}
\end{figure*}

The resulting locus for the star on a HR diagram is shown in Figure
\ref{fig_hrd}.  For comparison we show evolutionary tracks for massive
stellar models at solar metallicity.  We first consider models without
rotation and with rotation $v_{\rm ini} = 300$ \kms, shown as dotted
and dashed lines (respectively) in Figure \ref{fig_hrd}
(\citealt{hirschi04}; hereafter referred to as the Geneva group
tracks).  Stellar rotation can have a profound effect on the inferred
nature of the massive SN progenitors and on various observed
characteristics of the explosion, particularly at higher masses,
specifically, 15--25 M$_{\sun}$.  The Geneva group tracks have been
computed in steps of 3--5 M$_{\sun}$. The locus for the progenitor
candidate lies in between two sets of these models. A determination of
the initial mass for the progenitor candidate therefore requires
interpolation between these mass steps. Our ``by eye'' estimate is an
initial mass of $18 \pm 2\ {\rm M}_{\sun}$.

However, although the star appears to have a luminosity comparable to
the RSG stage of the Geneva group tracks, the fact that it appears to
be a yellow supergiant would require that it had undergone a ``blue
loop'' away from the RSG branch. A blue loop could result from
instabilities which occur in between the core He-burning and the 
shell H-burning stages, due to, for example, increasing H abundance,
contraction of the H envelope as a consequence of an expansion of the
He core, or a thermal non-equilibrium response of the stellar envelope
to the variation of the core luminosity \citep[][and references
  therein]{kippenhahn94,xu04}. However, in the Geneva group tracks a
late ``blue" excursion is missing at masses between 12 and
20\ M$_{\sun}$, since these models still have most of their H-rich
envelope at the termination of their calculation. The lack of a clear
blue loop entails only a slight deviation from the Hayashi track, and
this may sometimes occur only during the central C-burning phase
(R. Hirschi and S. Ekstr\"{o}m, 2009, private communication). For the
Geneva group models with rotation, blue loops may occur for masses
well above 15\ M$_{\sun}$ and below the limit for stars evolving to
the Wolf-Rayet phase. For this reason, we also consider the
\citet{bressan93} evolutionary tracks (hereafter referred to as the
Padova group tracks). For these tracks, the blue loop is present at
lower initial masses, such as 15 M$_{\sun}$ (the {\it solid line} in
Figure \ref{fig_hrd}). However, we note that these models do not
include rotation. Comparing the star's intrinsic properties with the
Padova group track, the star's initial mass is consistent with $15 \pm
2\ {\rm M}_{\sun}$. Regardless of the evolutionary models, the biggest
contributions to the uncertainty in this mass estimate come from the
photometric uncertainties for such a relatively faint star, the
uncertainties in the extinction, and the uncertainties in the
bolometric correction and effective temperature for a yellow
supergiant.

As can be seen, the inferred initial mass of the progenitor candidate
for SN~2008cn, based on the available evolutionary models, is within
the range of progenitor initial masses found so far for SNe II-P
(\citealt{smartt09}), albeit on the high side of that
range. Considering that SN~2008cn appears to be a member of the
high-luminosity SN subclass (see \S \ref{specph}), this estimate of
its progenitor's initial mass adds a valuable additional statistic,
previously only provided, possibly, by SN~2004et \citep{li05}.

Of course, until now we have only been considering the progenitor
candidate as a single star.  The output from HSTphot, in addition to
source position, flight-system magnitudes, and uncertainties, also
contains the ``sharpness'' of the identified objects (see
\citealt{dolphin00a} for further details), which indicates the
reliability that a detected source is indeed point-like. As
\citet{leonard08} recommend, a ``good star'' can be considered one
where the value of this parameter is between $-$0.3 and $+$0.3. From
our runs of HSTphot on both F555W and F814W images, the overall
sharpness of the progenitor candidate's profiles fell within these
recommended limits, $-$0.036 in F555W and $-$0.184 in
F814W. Additionally, the results also indicate that the ``object
type'' flag is ``1,'' further indicating that the source is likely
stellar (see \citealt{dolphin00a}). However, at the distance of the
host galaxy, 33.3 Mpc (\citealt{newman99}), a single {\sl HST}/WFPC2
pixel (at $0{\farcs}05$ pix$^{-1}$) is $\sim 8$ pc. Within this space
a large number of stars could be contained, possibly even an
unresolved stellar cluster. Although our progenitor candidate is quite
luminous in $V$ and $I$, we can safely discount the progenitor
candidate as a compact cluster, based on luminosity arguments
\citep[see][]{bastian05,crockett08}.

\begin{figure*}
\centering
\includegraphics[height=5truein,width=3.8truein,angle=-90]{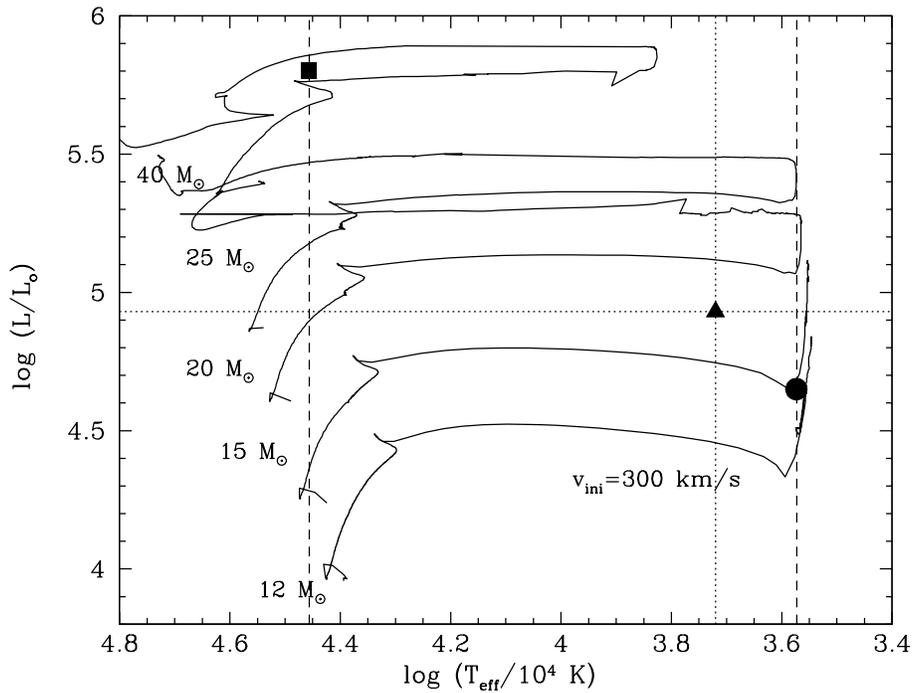}
\caption{A Hertzsprung-Russell diagram showing an example of a blue
  (B0\,I spectral type, {\it filled square}) and a red (M1\,I spectral
  type, {\it filled circle}) supergiant, which, when combined, result
  in the luminosity $L_{\rm bol}$ and $T_{\rm eff}$ ({\it dotted lines\/})
  estimated for the progenitor candidate of SN~2008cn 
   ({\it filled triangle}).  For comparison, 
  evolutionary tracks with $Z = 0.02$
  and $v_{\rm ini}=300$ \kms\ ({\it solid lines}; \citealt{hirschi04})
  are shown. The vertical {\it dashed lines\/} indicate the $T_{\rm
    eff}$ for the input blue and red supergiants.
\label{fig_redandblue}}
\end{figure*}

The yellow color of the SN progenitor could be the result of the
blending of two or more stars into a single stellar profile.  We
attempted to deconvolve the yellow supergiant as a possible unresolved
red and blue star combination, performing synthetic photometry using
the package {\it synphot\/} within IRAF/STSDAS of model stellar
spectra for blue supergiants (O8--B2 spectral types) and red
supergiants (K3--M4 spectral types).  The adopted values of effective
temperature, $V-I$ color, and bolometric correction ascribed to these
spectral types were obtained from \citet{levesque05},
\citet{humphreys84}, and \citet{kurucz93}. In Figure
\ref{fig_redandblue} we show an example of one of these combinations,
in comparison with the Geneva group tracks with rotation: the
superposition of a $M_{\rm ini} \approx 40\ {\rm M}_{\sun}$ B0\,I star
and a $M_{\rm ini} \approx 15\ {\rm M}_{\sun}$ M1\,I star. From such a
combination we can reproduce the intrinsic properties of the ``yellow"
star. Thus, a red supergiant with an initial mass of 15\ M$_{\sun}$
could be ``hidden" within the superposition of the light from one or
more brighter, bluer stars. Such a red star, in principle, could
instead be the progenitor of SN~2008cn.

Furthermore, \citet{prieto08} have found two eclipsing binaries containing massive yellow
supergiants, one in the Small Magellanic Cloud and one in Holmberg
IX. These authors point out that such an interacting binary could slow
down the evolutionary transition from blue to red supergiant for the
stars in the system, allowing the primary star to explode as a SN
while still yellow. They offer these binaries as possible analogs for
the progenitors of SNe 2004et and 2006ov. In a similar vein, it is
possible that the yellow progenitor candidate for SN~2008cn is a
similar close, massive binary system. Both the brightness and color of
the binary in Holmberg IX \citep{prieto08} are remarkably similar to
those of the SN 2008cn progenitor candidate.

\begin{figure*}
\centering
\includegraphics[height=5truein,width=3.8truein,angle=-90]{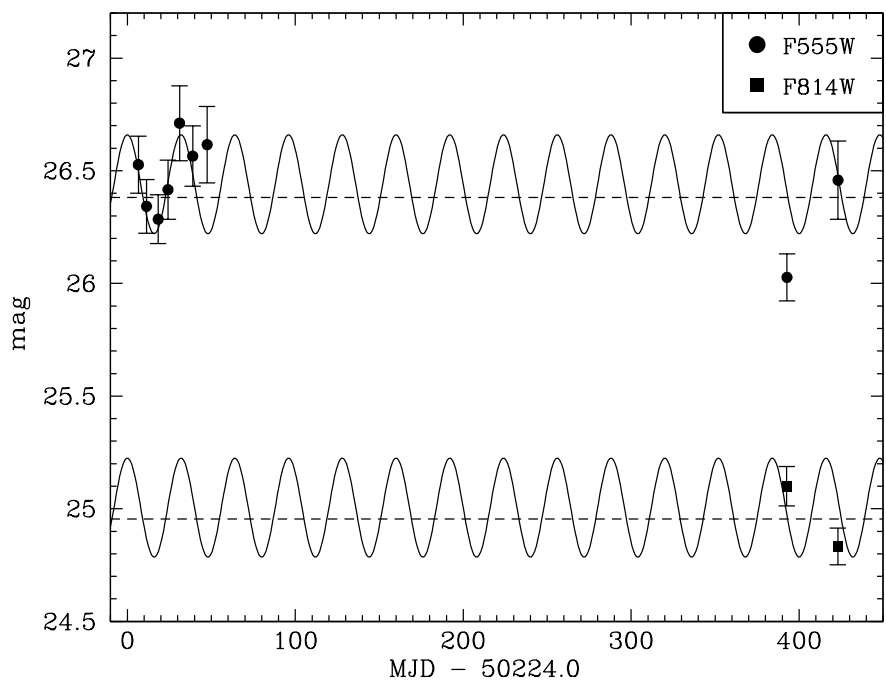}
\caption{{\sl HST\/} F555W ($\sim V$; {\it filled circles}) and F814W
  ($\sim I$; {\it filled squares}) magnitude vs. epoch for the
  progenitor candidate of SN~2008cn. Each point is the average
  magnitude at each epoch in each band. {\it Dashed lines} indicate
  the average magnitude in each band.  Model sinusoidal curves (simply
  representing an eclipsing binary system with components of equal
  brightness, {\it solid lines}) is also
  shown. The ``fits'' for the model curves were made purely by
  eye. See the text for discussion. Noticeable deviations from the
  model exist at some epochs.
\label{fig_varia}}
\end{figure*}

To that end, we are fortunate in the case of SN 2008cn that, for the
first time in the study of the progenitor of a SN II-P, pre-explosion
images exist at various epochs in two bands (\citealt{cohen95}
investigated the variability of the progenitor of the Type IIb 1993J
in M81 and did not find any). We can therefore search for 
variability in the brightness of the progenitor candidate. In Figure
\ref{fig_varia}, we show as a function of time the flight-system F555W
and F814W magnitudes, which we obtained by running HSTphot on all
CR-SPLIT pairs in each band at each epoch. As one can see, the
progenitor candidate {\it may\/} marginally show variation in
brightness in each band.  Such variability would have occurred within
11--12 yr prior to explosion, when the pre-SN observations were
made. In Figure \ref{fig_varia} we show for comparison model eclipses
(simple sinusoids) in each band with period $\sim$64 days and
amplitude 0.22 mag ({\it solid curves}), similar in shape to the
eclipses for the binary in Holmberg IX \citep{prieto08}. By eye, this set
of model light curves provides a reasonable representation of the
apparent variability. We also note that, from a simple $\chi^2$ analysis, we actually find a
smaller distribution in the residuals between data points and the
average magnitude in each band for the progenitor candidate 
({\it dashed lines} in Figure \ref{fig_varia}) than for the simple varying model. This is not surprising, since the star is
faint (average $V = 26.41$ mag) and the photometric uncertainties are
large (the star is detected in each of the CR-SPLIT pairs at only 
3--4$\sigma$).

Finally, we also consider the ramifications if the candidate is {\it
  not\/} the progenitor of SN 2008cn. If the progenitor has not been
detected in the pre-SN {\sl HST\/} images, the upper limits in each
band provide limits on the brightness, color, and initial mass of any
undetected star. As stated above, no other point sources could be
isolated in the pre-explosion {\sl HST\/} images within a
3$\sigma$-radius ($\sim$ 5 pixels) circle around the progenitor
candidate position. We therefore input artificial stars with $ 27
\lesssim V \lesssim 29$ mag (i.e., with brightnesses fainter than the
progenitor candidate) and colors $1.5 \lesssim V-I\ \lesssim 2.8$ mag
for an assumed RSG at random positions within a radius of 10 pixels
around the candidate's position using option 64 within HSTphot (see
the resulting upper limits in Table \ref{table_ph}). Applying the same
total extinction as assumed for the yellow supergiant progenitor
candidate and a corresponding bolometric correction of $-$1.25 mag
appropriate for RSGs (\citealt{levesque05}), we find an upper limit to
an undetected progenitor's bolometric luminosity of $L_{\rm bol}
\lesssim 10^{(4.65 \pm 0.23)}\ {\rm L}_{\sun}$. To convert this luminosity
limit to an upper limit on initial mass, we considered the
luminosities of stellar models with rotation at solar metallicity, in
the mass range 9--25 M$_{\sun}$ (\citealt{hirschi04}), at both the
endpoints of the models and at the end of core He burning (see
Smartt et al. 2008).

\begin{figure*}
\epsscale{1.0}
\centering
\includegraphics[height=5truein,width=3.8truein,angle=-90]{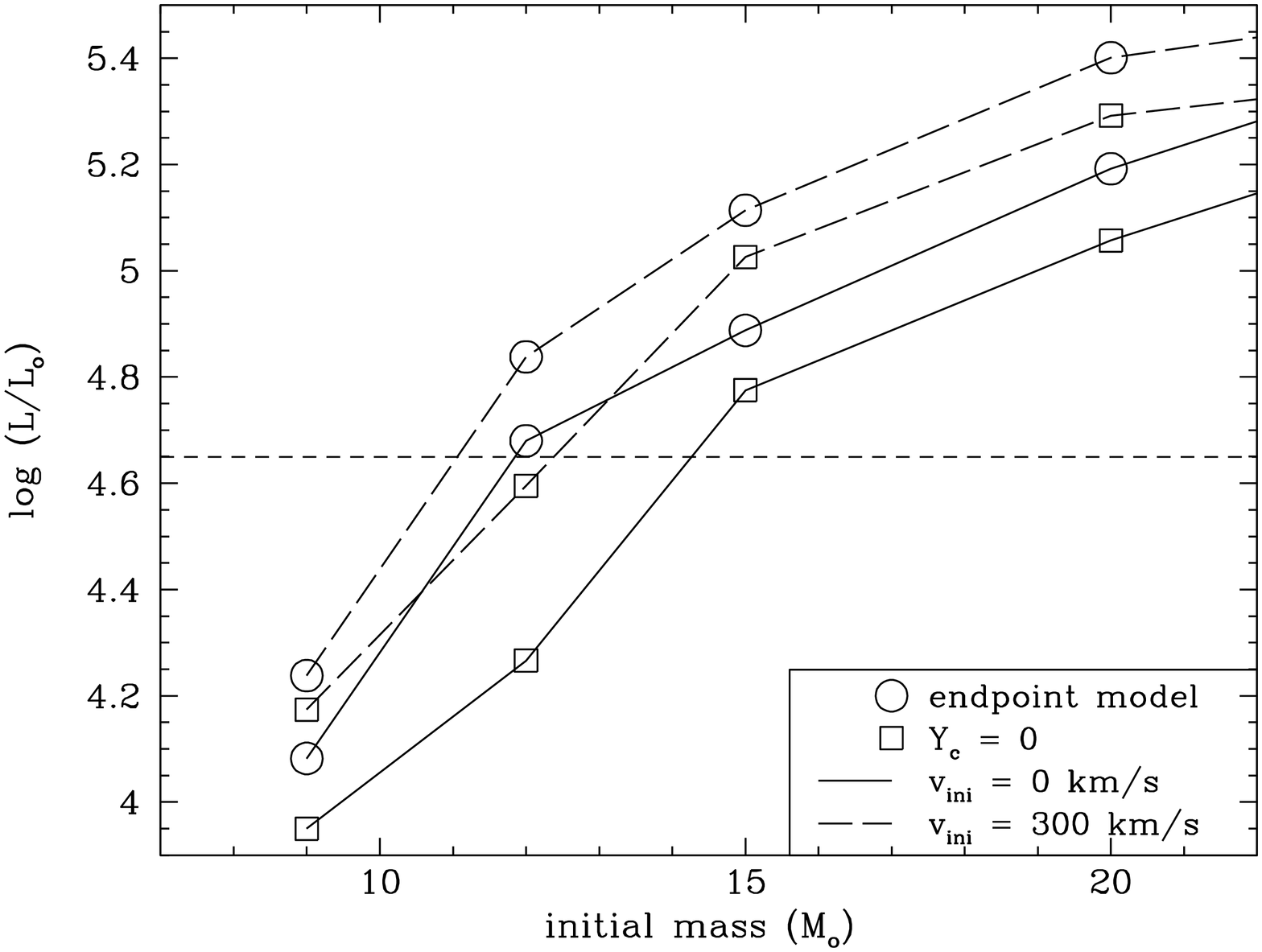}
\caption{Initial mass vs. endpoint luminosity of the stellar models
  (\citealt{hirschi04}) for $Z = 0.02$ at $v_{\rm ini} = 0$ \kms 
  ({\it solid lines}) and $v_{\rm ini} = 300$ \kms ({\it dashed lines}). 
  For each mass the luminosities corresponding to the
  endpoint of the model ({\it open circles}) and the end of the core
  He-burning phase ({\it open squares}) are shown.  The 
  {\it short-dashed line\/} indicates the upper luminosity limit for any
  red supergiant (in the range $ 27 \lesssim V \lesssim 29$ mag, $1.5
  \lesssim V-I \lesssim 2.8$ mag) undetected near the site of
  SN~2008cn.
\label{fig_limit}}
\end{figure*}

The results are shown in Figure \ref{fig_limit}. The upper limit to
any undetected RSG progenitor's luminosity, in the extreme case in
which such a star is at the end of the core He burning and without
rotation at explosion, implies that it would have $M_{\rm ini}
\lesssim 14\ {\rm M}_{\sun}$.  Considering, instead, what we believe
is a physically more realistic scenario of explosion at the model
endpoint luminosity for a rotating star, an undetected progenitor
would have $M_{\rm ini} \lesssim 11\ {\rm M}_{\sun}$. We note that
both limits are also consistent with the results for other SNe~II-P
(\citealt{smartt09}). However, given the correlations between the main
physical parameters for SNe II-P (see, e.g.,
\citealt{hamuy03,pastorello03,nadyozhin03}), we might expect that
high-luminosity SNe produce massive ejecta with high nickel masses 
($M_{\rm Ni}$) and, consequently, have high-mass progenitors. For
example, \citet{pastorello03}, using a semiÐanalytical model
\citep{zampieri03}, estimated that the progenitor of SN~1996W had an
initial mass of $15.2^{+6.4}_{-3.2}\ {\rm M}_{\sun}$. Since SN~2008cn
also appears to be of the high-luminosity subclass, we might expect
the $M_{\rm Ni}$ from this object, and therefore its progenitor mass,
to also be relatively high (unfortunately, given the limited data
available to us for SN~2008cn, we cannot estimate the actual $M_{\rm
  Ni}$ here). Such expected high masses for the high-luminosity SN
II-P progenitors stand at odds with the relatively low upper limits
that we have derived for an undetected progenitor of SN 2008cn.

That we cannot definitively identify the yellow supergiant candidate
as the SN~2008cn progenitor most likely results from the fact that
this is the most distant SN II-P (at 33.3 Mpc) for which anyone has
attempted the identification of a progenitor in pre-SN archival
images.  In fact, it is only due to the extraordinarily deep
pre-explosion {\sl HST\/} archival data that we are even able to
attempt this. To explore the effects of the distance further, we
estimated a limit in distance beyond which we can no longer
confidently identify single stars in {\sl HST\/} images at the typical
spatial resolution of the WFC chip (it is relatively rare that a SN
site will be located in archival PC chip images, due to that chip's
much smaller area).  We note that we expect the massive, RSG
progenitors of SNe II-P to be found in relatively crowded environments
with other massive stars --- this, in fact, is generally borne out
from detailed studies of these progenitors (see \citealt{smartt09},
and references therein).  We conducted a simple test using the {\sl
  HST\/} images containing the progenitor of the comparatively nearby
and well-studied SN II-P 2003gd (\citealt{vandyk03}).  SN~2003gd, in
the host galaxy M74 at 7.2 Mpc, was most likely the explosion of a RSG
with initial mass $\sim 8\ {\rm M}_{\sun}$, identified by
\citet{vandyk03} as ``Star A'' in an archival WFPC2 WF2 chip image in
F606W (and later confirmed by \citealt{smartt04}).  Nearby ($\sim
10.5$ pc) to Star A is a bluer object, ``Star B.''  We artificially
degraded the spatial resolution of the SN~2003gd pre-explosion image
until it was no longer possible to distinguish between the two
individual stars in the image.  The resulting degraded resolution
corresponds to the equivalent resolution of an image of a galaxy at a
distance of $\sim 20$ Mpc.  We conclude, therefore, that for galaxies
which host SNe II-P at distances $\gtrsim 20$ Mpc, such as NGC~4603,
it becomes far more difficult to identify unambiguously their (RSG)
progenitors. This distance sets a practical limit on using even the
high spatial resolution of {\sl HST\/} images to address the problem
of progenitor identification.



\section{Conclusions}\label{conclusions}

We have shown, based on a limited dataset, that the SN II-P 2008cn in
NGC~4603 appears to be of the high-luminosity subclass, similar to
SN~1996W \citep{pastorello03}, SN~2003hn \citep{krisciunas09}, and
possibly SN 2004et \citep{sahu06}.  Its $V$-band luminosity some weeks
after explosion is $\sim$0.5 mag brighter than that of, for
example, the normal SN II-P~1999em \citep{hamuy01,leonard02}. The
progenitors of normal SNe II-P have all been determined or inferred to
be RSGs \citep{smartt09}.  It is unclear at this point, due to a
relative lack of examples, whether high-luminosity SNe II-P, such as
SN 2008cn, also arise from RSGs.

High-precision relative astrometry has been employed to identify a
candidate progenitor for SN 2008cn, based on a comparison of high
spatial resolution {\sl HST\/} imaging both of the SN at late times
and of the host galaxy containing the SN site prior to
explosion. Comparison of the difference between the SN and candidate
positions with the astrometric uncertainties suggests that we may well
have identified the SN progenitor, although this is not entirely
certain.  In estimating an initial mass for the progenitor candidate,
we have been careful to transform the observed photometry into
intrinsic stellar quantities, specifically the bolometric luminosity
and the effective temperature of the star.  We have also considered
for comparison the most recent available stellar evolutionary models
for massive stars, including those with rotation (in particular, the
Geneva group models; the Padova group models do not include
rotation). From this we have estimated an initial mass for the
progenitor of SN~2008cn of $15 \pm 2\ {\rm M}_{\sun}$, which is within
the initial mass range for SN II-P progenitors, based on all previous
attempts that have been made to identify the star \citep{smartt09}.

However, the star's color is more yellow than we would expect.  This
unusual color could arise from the star exploding while experiencing
an evolutionary blue loop, which is theoretically not expected to
occur. Alternatively, the stellar profile could consist of two or more
stars in a blend, with a possible combination of one or more blue and
red supergiants resulting in the yellow color. In this case, a blended
stellar profile could ``conceal," for example, a $\sim 15\ {\rm
  M}_{\sun}$ RSG, which could nominally be the
progenitor. Additionally, the blend could be a massive, interacting
binary, analogous to the yellow supergiants recently found in
extragalactic eclipsing binary systems \citep{prieto08}. That SN
2008cn, like SN 2004et, is a high-luminosity SN II-P makes this a
particularly intriguing possibility. The progenitor of SN 2004et has
been identified as a yellow supergiant \citep{li05}, also
coincidentally with $M_{\rm ini} \approx 15\ {\rm M}_{\sun}$, although
this identification is controversial (see \citealt{smartt09}). Close
interaction in such a massive binary system could prevent the primary
from evolving to the RSG phase, allowing the star to explode as a
yellow supergiant.  SN~2008cn could well represent an additional
example of what is likely a relatively rare event.

Establishing the precise position of the faint progenitor candidate in
the pre-explosion archival images has proven to be challenging,
especially with the presence of the two brighter stars near the
fainter candidate and the relatively large distance to the host
galaxy. The slight excess in the positional difference in declination,
relative to the positional errors, prevents us from definitively
assigning this star as the progenitor. If the candidate yellow
supergiant is {\it not\/} the SN~2008cn progenitor, then the upper
limits to detecting any star within $\sim$10 pixels ($\sim 80$ pc) of
this candidate result in a conservative upper limit to the initial
mass of a RSG progenitor of $M_{\rm ini} \lesssim 14\ {\rm
  M}_{\sun}$. Ultimately, the best test of the nature of the SN~2008cn
progenitor is to observe this field again with {\sl HST\/} (or the
{\sl James Webb Space Telescope}) several years later, when the SN has
faded to nondetectability. The definitive indication would be if the
yellow star has vanished, or is significantly diminished in
brightness, or has survived intact.


\acknowledgments

Financial support for this work was provided by NASA through grants
AR-10952, AR-11248, and GO-11119 from the Space Telescope Science
Institute, which is operated by Associated Universities for Research
in Astronomy, Inc., under NASA contract NAS 5-26555. We acknowledge
the National Science Foundation (NSF) through grant AST--0306969 for
support of the Carnegie Supernova Program.  N.E.R. thanks Andrew
Dolphin for his help with HSTphot, Philip Massey and Raphael Hirschi
for their help with the supergiant parameters and models, Artemio
Herrero for stimulating discussions, and Stefano Benetti for his help
in the determination of SN parameters. M.H. acknowledge support from
FONDECYT through grant 1060808, the Millennium Center for Supernova
Science through grant P06-045-F funded by ``Programa Bicentenario de
Ciencia y Tecnolog\'ia de CONICYT'' and ``Programa Iniciativa
Cient\'ifica Milenio de MIDEPLAN,'' Centro de Astrof\'\i sica FONDAP
1501000, and Center of Excellence in Astrophysics and Associated
Technologies (PFB 06). This publication makes use of data products
from the Two Micron All Sky Survey, which is a joint project of the
University of Massachusetts and the Infrared Processing and Analysis
Center/California Institute of Technology, funded by NASA and the NSF.

{\it Facilities:} \facility{HST (WFPC2)}, \facility{Magellan:Baade
(IMACS)}, \facility{Swope}


\clearpage

\clearpage

\clearpage

\begin{deluxetable}{llllcl}
\tablewidth{0pt}
\tabletypesize{\scriptsize}
\tablecaption{Imaging Data for the SN~2008cn Site\label{table_observ}}
\tablehead{
\colhead{UT Date}           & \colhead{Telescope/Instrument}      &
\colhead{Filter}          & \colhead{Exp. Time}  &
\colhead{\# Images or}  & \colhead{{\sl HST\/} Image}
\\
\colhead{}           & \colhead{}      &
\colhead{}          & \colhead{(s)}  &
\colhead{CR-SPLIT Pairs}          & \colhead{Sets}
}
\startdata
\bf{Pre-explosion:}\\
1996 May 27 & {\sl HST}/WPFC2 & F555W & 900/1300 & 3 & u38l010[1--6]t \\
1996 June 01 & {\sl HST}/WPFC2 & F555W & 900/1300 & 3 & u38l020[1--6]t\\
1996 June 08 & {\sl HST}/WPFC2 & F555W & 900/1300 & 3 & u38l030[1--6]t\\
1996 June 14 & {\sl HST}/WPFC2 & F555W & 900/1300 & 3 & u38l040[1--6]t \\
1996 June 20 & {\sl HST}/WPFC2 & F555W & 900/1300 & 3 & u38l050[1--6]t\\
1996 June 28 & {\sl HST}/WPFC2 & F555W & 900/1300 & 3 & u38l060[1--6]f\\
1996 July 07 & {\sl HST}/WPFC2 & F555W & 900/1300 & 2 & u38l070[1--4]f\\
1997 June 17 & {\sl HST}/WPFC2 & F555W & 1100/1300 & 2 & u38l080[1--4]f\\
1997 June 17 & {\sl HST}/WPFC2 & F814W & 1100/1300 & 3 & u3zc010[1--6]m\\
1997 July 17 & {\sl HST}/WPFC2 & F814W & 1100/1300 & 3 & u3zc020[1--6]m\\
1997 July 18 & {\sl HST}/WPFC2 & F555W & 1100/1300 & 2 & u38l090[1--4]f\\
\tableline
\\[-6pt]
\bf{Post-explosion:}\\
2008 June 1 & Swope/CCD & $V$ &  100 & 1 & \nodata \\
2008 June 15 & MagI/IMACS & $V$ &  100 & 2 & \nodata \\
2008 August 26 & {\sl HST}/WPFC2\tablenotemark{a} & F555W & 1100/1300 & 7 &  ua22030[1--7]m \\
2008 August 26 & {\sl HST}/WPFC2\tablenotemark{a} & F814W & 1100/1300 & 8 & ua220308-9/a-fm\\
2009 July 01 & CTIO/ANDICAM & $VRI$ &  360 & 3/2/2 & \nodata \\
\enddata
\tablecomments{{\sl HST}/WPFC2 = {\sl Hubble Space Telescope\/} + Wide
  Field Planetary Camera 2, WF2 chip, $0{\farcs}1$ pix$^{-1}$;
  Swope/CCD = 1.0-m Swope telescope + Direct CCD Camera,
  $0{\farcs}435$ pix$^{-1}$; MagI/IMACS = Baade 6.5-m Magellan
  Telescope + IMACS Long-Camera, $0{\farcs}2$ pix$^{-1}$; CTIO/ANDICAM
  = CTIO 1.3-m Telescope + ANDICAM Camera, $0{\farcs}369$ pix$^{-1}$.}
\tablenotetext{a}{In these observations SN~2008cn is located on
the PC chip, $0{\farcs}046$ pix$^{-1}$.}
\end{deluxetable}

\begin{deluxetable}{lcc}
\tablewidth{0pt}
\tablecaption{Differences Between the SN and Progenitor Candidate 
Positions\label{table_error}}
\tablehead{
\colhead{}           & \colhead{$V$ ($\alpha/\delta$)}      &
\colhead{$I$ ($\alpha/\delta$)}
}
\startdata
Uncertainty in the progenitor position (mas) & 7/3 & 0/2\\
Uncertainty in the SN position (mas) & 2/1 & 4/0\\
Geometric transformation (mas) & 29/28 & 29/21\\
Total uncertainty (mas) & 30/28 & 29/21\\
\tableline
Difference in position (mas) & 9/40 & 5/33\\
\enddata
\tablecomments{Uncertainties in the SN and candidate right ascension,
  $\alpha$, and declination, $\delta$, in milliarcsec, were estimated
  as the standard deviation of the average. Geometric transformation
  errors are derived from the positional differences between the
  fiducial stars used in the transformation. The total uncertainty is
  the quadrature sum of these uncertainties. The last line lists the
  residual difference between the SN and progenitor position after the
  geometric transformation.}
\end{deluxetable}

\begin{deluxetable}{lccccc}
\tablewidth{0pt}
\tablecaption{Photometry of SN~2008cn, the Progenitor Candidate,
and Artificial Stars \label{table_ph}} \tablehead{ \colhead{} &
\colhead{UT Date} & \colhead{F555W}    & \colhead{F814W}  &
\colhead{$V$}  & \colhead{$I$} \\
\colhead{} & \colhead{} & \colhead{(mag)}    & \colhead{(mag)}  &
\colhead{(mag)}  & \colhead{(mag)}} \startdata
SN~2008cn  & 2008 June 1 & \nodata & \nodata & 16.51$\pm$0.22 & \nodata \\
SN~2008cn  & 2008 June 15 & \nodata & \nodata & 16.46$\pm$0.02 & \nodata\\
SN~2008cn  & 2008 Aug. 26 & 18.15$\pm$0.01 & 16.64$\pm$0.01 & 18.13$\pm$0.04 & 16.60$\pm$0.05\\
SN~2008cn  & 2009 July 1 & \nodata & \nodata & $\gtrsim$21.0 & $\gtrsim$20.5 \\
\tableline
\\[-6pt]
Progenitor candidate & \nodata & 26.41$\pm$0.04  &  25.01$\pm$0.06  &  26.37$\pm$0.04  &  24.96$\pm$0.07  \\
Artificial star limit$^a$ & \nodata & $\gtrsim$28.1 & $\gtrsim$25.4 & $\gtrsim$28.0 & $\gtrsim$25.4 \\
\enddata
\tablenotetext{a}{The 2$\sigma$ upper limits to the magnitudes for the
  undetected artificial stars are the averages of the determinations
  from each set of CR-SPLIT pairs.  Uncertainties were estimated in
  the photometric calibration for the ground-based observations, and
  by HSTphot for the {\sl HST} images. }
\end{deluxetable}

\end{document}